\documentstyle[titlepage]{article}
\newcommand{\R}{{\rm I\kern-2pt R}}

\newtheorem{proposition}{Proposition}[section]

\newtheorem{theorem}{Theorem}[section]
\newtheorem{remark}{Remark}[section]
\newtheorem{definition}{Definition}[section]

\begin{document}
\begin{center}
{\bf The Minimal Polynomial of Some Matrices Via Quaternions\\} 
Viswanath Ramakrishna$^{\dagger}$ $\&$ Yassmin Ansari$^{\dagger}$, 
Fred M. Costa$^{\ast}$\\ 
$\dagger$ Department of Mathematical Sciences
and Center for Signals, Systems and Communications\\
University of Texas at Dallas\\
P. O. Box 830688\\
Richardson, TX 75083 USA\\
$\ast$ Keane Incorporated, 12000 Ford Rd\\
Farmer's Branch, TX 75234\\ 
email: vish@utdallas.edu 
\end{center}

\begin{abstract}
This work provides explicit characterizations and formulae for 
the minimal polynomials
of a wide variety of 
structured $4\times 4$ matrices. These include 
symmetric, Hamiltonian and orthogonal matrices. Applications
such as the complete determination of the Jordan structure
of skew-Hamiltonian matrices and the computation of
the Cayley transform are given. 
Some new classes of matrices are uncovered,
whose behaviour insofar as minimal polynomials are concerned, is remarkably
similar to those of skew-Hamiltonian and Hamiltonian matrices.
The main technique is the invocation of
the associative algebra isomorphism between the tensor product
of the quaternions with themselves and the algebra of real
$4\times 4$ matrices.

\end{abstract}

\section{Introduction}

The minimal polynomial of a matrix is the unique monic polynomial
of minimal degree which annihilates the matrix. It has several
theoretical and practical uses. It provides information about the
Jordan structure 
of the matrix, and in some situations can nearly determine it. 
Its principal utility, arguably, 
is in computing functions of a matrix such as the matrix
exponential and the Cayley transform. While any annihilating polynomial
can be used for this purpose, the complexity of the resultant expression
is naturally minimal when the minimal polynomial is used. 

If one knows the Jordan structure of the matrix then its minimal polynomial
is easily computed. However, since the former is difficult to arrive at,
this is rarely advisable. 
Essentially any mechanism
which {\it explicitly} detects linear dependence at the earliest stage in 
the matrices
$I, A, A^{2}, \ldots , A^{n}$ (in that order) will yield the minimal
polynomial, \cite{hhorni,grammoment}.
In this work we use quaternions to achieve the same for specific
structured $4\times 4$
matrices. Whilst, $4\times 4$ matrices are amenable to the techniques
of \cite{hhorni,grammoment}, 
the corresponding calculations can be quite difficult, and
would not produce the closed form expressions for minimal polynomials
presented herein.
More discussion on this issue is presented in Section 5.  
In the method
proposed here one replaces matrix calculations (specifically computing
$A^{k}$) via quaternion calculations. Not only does this simplify such
calculations, but it also yields elegant geometric interpretations of
situations wherein the minimal polynomial is a particular polynomial.
Of course, this methodology does not extend to higher dimensions
immediately (see, however, the discussion in Section 5), but
$4\times 4$ matrices already cover several important applications
to warrant the investigation of such a technique. 
In quantum computation, quantum optics,
computer graphics, robotics etc., much of the analysis is reducible
to the study of $4\times 4$ matrices [see, for instance,
\cite{goongi,fourporti,selig}].

The isomorphism of $H\otimes H$ with $M (4, R)$ is a central point
in the theory of Clifford algebras. 
So a natural question is whether Clifford algebra isomorphisms can be used
for similar purposes. In fact, the interesting work
of \cite{cliffminpolyi} uses the symbolic
and numerical computation 
the (real) minimal polynomial of matrices via their Clifford algebra
representatives, for exponentiation of matrices. The difference between
the work of \cite{cliffminpolyi} and the results here, insofar as the
problem of computation of minimal polynomials of matrices in 
$M(4, R)$ is concerned, is that the structural (i.e., ``geometrical")
conditions given here on the entries of a matrix for it to possess a given
minimal polynomial are missing in \cite{cliffminpolyi}. There are of course
other differences. Section 5 discusses this issue briefly.

It is appropriate at this point to record some history of the linear algebraic
applications of the
isomorphism between $H\otimes H$ and $M_{4}(R)$.
This isomorphism is central to the theory of Clifford algebras, \cite{pertii}.
However, it is only relatively recently been put into use for
linear algebraic purposes.
To the best of our knowledge, the first instance seems to be the work
of \cite{kyf}, where it was used in the study of linear maps preserving the
Ky-Fan norm. Then in \cite{haconi}, this connection was used
to obtain the Schur canonical form explicitly for real $4\times 4$ 
skew-symmetric matrices. Next, is the work of \cite{nii,ni,niii}, wherein
this connection was put to 
innovative use for solving eigenproblems 
of several classes of structured $4\times 4$ matrices.
In \cite{expistruc,expisufour}, this isomorphism was used to
explicitly calculate the exponentials of a wide variety
of $4\times 4$ matrices.
Finally, in \cite{noncompactportion} it was used to obtain, among other things,
the polar decomposition of $4\times 4$ symplectic matrices via the solution of
$2\times 2$ linear systems of equations.

The balance of this paper are organized as follows. In the next section
basic notation and preliminary facts are reviewed. The next section
contains all the main results on minimal polynomials obtained by our
method. Since many of the proofs are similar we provide proofs for 
only a part of the announced results. The fourth section contains three
applications. The first is to the complete determination of the Jordan
structure of $4\times 4$ skew-Hamiltonian matrices. The second illustrates
the usage of the results on minimal polynomials to calculate the 
Cayley transform in closed form. The final application is to the
determination of the singular values of $3\times 3$ real matrices.
The next section discusses extension of the results of section 3 via
the use of Clifford Algebras. In particular, classes
of matrices are uncovered which behave very similar to skew-Hamiltonian
matrices insofar as minimal polynomial matrices are concerned. Their block
structures do not suggest this similarity. This section also provides a brief 
comparison of our technique with that of \cite{cliffminpolyi,grammoment}. 
The final section offers some conclusions.

\section{Notation and Preliminary Observations}
                                                                               
The classes of real matrices discussed in this work are as follows:
\begin{itemize}
\item Skew-symmetric matrices, i.e., $X$ satisfying $X^{T} = -X$.
\item Hamiltonian matrices, i.e., matrices $H$ satisfying
$H^{T}J_{2n} = - J_{2n}H$, where $J_{2n} = \left (\begin{array}{cc}
0_{n} & I_{n}\\
-I_{n} & 0_{n}
\end{array} \right )$.
\item Perskewsymmetric matrices, i.e, matrices $X$ satsifying
$X^{T}R_{n} = -R_{n}X$, where $R_{n}$ is the $n\times n$ matrix
containing $1$s on its main anti-diagonal and $0$s elsewhere.
$R_{n}$ is sometimes denoted $F$ and is called
the flip matrix.
\item For a matrix $X\in M(n, R)$, we denote by $X_{F}$, its adjoint with
respect to the non-degenerate bilinear form defined by $R_{n}$,
i.e., it is the matrix $R_{n}X^{T}R_{n}$.

\item For a matrix $X\in M(2n, R)$, we denote by $X_{H}$, its adjoint with 
respect to the non-degenerate bilinear form defined by $J_{n}$,
i.e., it is the matrix $-J_{2n}X^{T}J_{2n}$.
\item Symmetric matrices, i.e., $X$ with $X^{T} =X$.
\item Skew-Hamiltonian matrices, i.e, matrices $X$ satisfying
$X^{T}J_{2n} = J_{2n}X$.
\item Special orthogonal matrices, i.e., $X$ with $X^{T}X = XX^{T}
= I_{n}$ and ${\mbox det}(X) = 1$.
\end{itemize}

These classes were picked because i) they are ubiquitous in applications;
and ii) in most cases, as will be seen subsequently, elegant geometric
conditions on their quaternionic representations can be given which
ensure their possessing a certain minimal polynomial. Matrices
such as persymmetric and symplectic matrices do not seem that amenable
by the latter consideration, and therefore are not considered here. We note,
however, that in the final section we discuss how quaternion techniques
can be used to {\it compute the minimal polynomial of a general matrix in
$M (4, R)$.}    

\begin{definition}

{\rm $H$ stands for the real division algebra of the quaternions.
${\mathcal P}$ stands for the purely imaginary quaternions.}

\end{definition}

\noindent We will tacitly identify an element of ${\mathcal P}$
with the corresponding vector in $R^{3}$. With this understood, 
the following two identities will be frequently used.
\begin{itemize}
\item Let $p, q\in {\mathcal P}$. Then $pq =
-(p.q)1 +p\times q$.
\item Let $p,q,r\in R^{3}$. Then
\begin{equation}
\label{vectortriple}
p\times (q\times r) = (p.r)q-(p.q)r
\end{equation}
\end{itemize}

\noindent {\bf $H\otimes H$ and $gl(4, R)$}: The algebra isomorphism
between $H\otimes H$ and $ M_{4}(R)$ (also denoted
by $gl(4, R)$), which is central to this work, may be summarized as follows:

\begin{itemize} 
\item Associate to each product
tensor $p\otimes q\in H\otimes H$,
the matrix, $M_{p\otimes q}$, of the map which sends $x\in H$
to $px\bar{q}$, identifying $R^{4}$ with $H$ via the basis $\{1,i,j,k\}$.
Here, $\bar{q} = q_{0} - q_{1}i - q_{2}j - q_{3}k$

\item Extend this to the full tensor product by linearity. 
This yields an associative algebra isomorphism
between $H\otimes H$ and $M_{4}( R)$. Furthermore, a basis for
$gl(4, R)$ is provided by the sixteen matrices $M_{e_{x}\otimes e_{y}}$
as $e_{x}, e_{y}$ run through $1, i, j, k$.
In particular, $R_{4}$, the matrix intervening in the definition of
perskewsymmetric matrices, and $J_{4}$, the matrix used in the
definition of Hamiltonian and skew-Hamiltonian matrices, represented
respectively, by $M_{j\otimes i}$ and $M_{1\otimes j}$, belong to 
this basis.

\end{itemize}

\vspace*{9mm}

\noindent {\bf Quaternion Representations of Special Classes of Matrices:}
Throughout this work, the following list of
$H\otimes H$ representations of the above classes of matrices
will be used:
\begin{itemize}
\item {\it Skew-Symmetric Matrices:} 
$s\otimes 1 + 1\otimes t$ with
$s, t\in {\mathcal P}$.
\item {\it Hamiltonian Matrices:}
$b(1\otimes j) +
p\otimes 1 + q\otimes i + r\otimes k$, with $b\in {\mathcal R}$ and
$p, q, r\in {\mathcal P}$.
\item {\it Perskewsymmetric Matrices:}
 $r\otimes i + j\otimes s + \alpha (1\otimes i) + \beta (j\otimes 1)$,
with $r, s \in {\mathcal P}$ and $\alpha, \beta \in {\mathcal R}$.
\item {\it Symmetric Matrices:} $a1\otimes 1 + p\otimes i +
q\otimes j + r\otimes k$, with $a\in {\mathcal R}$
and $p,q,r\in {\mathcal P}$.
\item {\it Skew-Hamiltonian Matrices:}
$b(1\otimes 1)
+ p\otimes j + 1\otimes (ci + dk)$, with $b, c, d\in {\mathcal R}$ and
$p\in {\mathcal P}$.
\item {\it Special Orthogonal Matrices:} 
$u\otimes v$, with $u, v$ unit quaternions, i.e., $\mid\mid u\mid\mid
= \mid\mid v\mid\mid = 1$.
\end{itemize}  
These can be easily obtained from the
entries of the $4\times 4$ matrix in question (see 
\cite{ni,nii,niii} for some instances). The key to this consists of
the following two observations
\begin{itemize}
\item Conjugation in $H\otimes H$ corresponds to matrix transposition  
in $gl(4, R)$, i.e., $M_{\bar{p}\otimes \bar{q}} = 
(M_{p\otimes q})^{T}$.
This is why, for instance,
symmetric matrices correspond to $c(1\otimes 1) + p\otimes i
+ q\otimes j + r\otimes k$ with $p,q,r$ purely imaginary, and skew-symmetric
matrices correspond to $s\otimes 1 + 1\otimes t$, $s,t\in P$.
\item Hamiltonian (resp. skew-Hamiltonian) matrices are expressible as 
$J_{2n}S$, with $S$ symmetric (resp. skew-symmetric). Similarly
persymmetric (resp. perskew-symmetric) matrices are expressible as
$R_{n}S$ with $S$ symmetric (resp. skew-symmetric).
Thus, for instance, perskewsymmetric matrices are represented by
$(j\otimes i)[s\otimes 1 + 1\otimes t]$, $s,t\in {\mathcal P}$. This simplifies
to $p\otimes i + \alpha (j\otimes 1) + j\otimes q
+ \beta (1\otimes i)$ with $p\in {\mbox span} \ \{i,k \},
q \in {\mbox span} \ \{j,k\}, \alpha , \beta \in R$. If such a matrix
is simultaneously symmetric, then $\alpha = \beta = 0$, etc.,
\end{itemize}
Combining these two observations with the explicit forms of the sixteen
matrices, $M_{e_{x}\otimes e_{y}}$  
leads to $H\otimes H$ representations, in terms of the entries of the matrices.
For the first five classes, the expressions for 
the $H\otimes H$ representations are linear in the entries of the matrix.
See \cite{ni,nii,niii} for these expressions.
For special orthogonal matrices, the entries of the matrix are quadratic
in $u$ and $v$.
See \cite{fourporti} for an algorithmic determination
of the unit 
quaternions $u$ and $v$ from the entries of a special orthogonal matrix.  

By way of illustration,
the requisite expression for the quaternionic representation  
for a skew-Hamiltonian matrix is  provided below.

\begin{proposition}\cite{nii}
\label{skewHamquartrepnI}
{\rm Let $X$ be a skew-Hamiltonian matrix. Its $H\otimes H$ representation
is $b(1\otimes 1)
+ p\otimes j + 1\otimes (ci + dk)$, with $b, c,d \in R$ and $p$ a purely
imaginary quaternion. The formulae relating these to $X$'s 
entries are as follows:  
i) $b = \frac{1}{2}(X_{11} + X_{22})$;
ii) $ p = \frac{1}{2} [ (X_{32} - X_{14}), (X_{11} - X_{22}),
(X_{12} + X_{21})]$;  iii) $c = \frac{1}{2}(X_{12} - X_{21})$;
iv) $d = \frac{1}{2} (X_{14} + X_{32})$.}
\end{proposition}
  
We close this section with the notion of the reverse of a polynomial.

\begin{definition}
{\rm If $p(x) = \sum_{i=0}^{n}a_{i}x^{i}$ is a polynomial of degree $n$,
then its reverse is the polynomial $ p_{{\mbox rev}}(x) =
\sum_{i=0}^{n}a_{n-i}x^{i}$.}
\end{definition}
 
\section {Minimal Polynomials of Classes of $4\times 4$ Matrices}
We begin with a simple proposition, applicable in arbitrary dimensions, 
which reduces the list of possible minimal
polynomials for some of the matrices to be considered here.

\begin{proposition}
\label{shortlist}
{\rm
\begin{itemize} 
\item I) Let $A^{T}$ be similar to $-A$. If the degree of the minimal
polynomial of $A$ is even, then its minimal polynomial is an even polynomial. 
If the degree of the minimal polyomial is odd, then it is an odd polynomial. 
\item II) Let $A^{-1}$ be similar to $A^{T}$. Then the constant term
in its minimal polynomial is either $+1$ or $-1$. If it is the former,
then its minimal polynomial equals its reverse. If it is the latter it
is minus its reverse.
\end{itemize}
}
\end{proposition}

\noindent {\it Proof:} I) Let $q_{A}(x)
= x^{k} +\sum_{i=0}^{k-1}a_{i}x^{i}$ be the minimal polynomial of $A$
(and thus of $A^{T}$).
Then clearly the polynomial $p(x) = x^{k} + \sum_{i=0}^{k-1}(-1)^{i}
a_{i}x^{i}$ annihilates the matrix $-A$.  So, $p(x)$, which is monic, has to 
be the
minimal polynomial of $-A$. Indeed, if $q_{(-A)}(x) = x^{l}
+\sum_{i=0}^{l-1}b_{i}x^{i}$, was the minimal polynomial of $-A$
(with $l < k$) then $x^{l} + \sum_{i=0}^{l-1}(-1)^{i}
b_{i}x^{i}$ annihilates $A$, contradicting the minimality of $q_{A}(x)$. 
Thus, since $-A$ and $A^{T}$ are
similar, $q_{A}(x) = p(x)$, and the result follows.

II) Let $q_{A}(x)
= x^{k} +\sum_{i=0}^{k-1}a_{i}x^{i}$ be the minimal polynomial of $A$.
Since $A$ is invertible $a_{0}\neq 0$.
Then, we find that
$p(x) =\frac{1}{a_{0}}( x^{k} + 
\sum_{i=0}^{k-1}a_{k-i}x^{i})$ annihilates $A^{-1}$.
Now, by an argument similar to I),  $p(x)$ has to be the minimal polynomial
of $A^{-1}$. The similarity of $A^{T}$ and $A^{-1}$ then implies
the equality of $p(x)$ and $q_{A}(x)$. This forces $a_{0}$ to be either
$1$ or $-1$. 
This then means $p(x)$ equals its reverse when $a_{0} = 1$ or it
equals minus its reverse when $a_{0} = -1$, and the result follows.  
$\diamondsuit$ 

\begin{remark}
{\rm 
Note that even if ${\mbox det}(A) = 1$, for matrices as in II) of
Proposition (\ref{shortlist}), it is possible that $a_{0} =-1$, and thus
the minimal polynomial equals minus its reverse. This is in sharp contrast
with the characteristic polynomial of such an $A$, which always equals its
reverse.}
\end{remark}  

Proposition (\ref{shortlist}) shortens
the list of possible minimal polynomials for
skew-symmetric, Hamiltonian, perskewsymmetric and special orthogonal matrices,
since in each of these cases $A^{T}$ is similar to either
$-A$ or $A^{-1}$. 
When $A^{T}$ is similar to $A$, there are no such general results.

Next follow our main results about minimal polynomials. As mentioned
in Section 1, we detail only those 
cases where one has an ``elegant" condition on the $H\otimes H$
representations of the matrix in question which is equivalent to the
matrix having the said polynomial as its minimal polynomial. This already
contains an extensive collection of useful matrices.   
Furthermore, since the proofs are similar, we present details only
for some cases.
\begin{theorem}
\label{skewsymm}
Minimal Polynomials of Antisymmetric Matrices:
{\rm Let $S$ be antisymmetric, with representation $s\otimes 1 + 1\otimes t$.
Then,
\begin{itemize}
\item $S$ has a quadratic minimal polynomial, which equals 
$x^{2} + \lambda^{2}$, iff
precisely one of $s$ or $t$ is equal to zero. 
Furthermore, in this case, $\lambda^{2}$ is either $s.s$ or  $t.t$.  
\item $S$ has a cubic minimal polynomial, which equals 
$p(x) = x^{3} + (\lambda^{2} + 2l)x$,
iff 
\[
\mid\mid s\mid\mid^{2} = \mid\mid t \mid\mid^{2} \neq 0 
\]
Furthermore, in this case, $\lambda^{2} = s.s + t.t$ and  $l= 
\mid\mid s\mid\mid^{2}$. 
\item If none of the above conditions hold, the minimal
polynomial is the  characteristic polynomial which equals 
$p(x) = x^{4} + 2\lambda^{2}x^{2} - (4l^{2} 
- \lambda^{4})$, with $\lambda$ and $l$ as above.     
\end{itemize}
}
\end{theorem} 

\begin{theorem}
\label{Hamilt}
Minimal Polynomials of Hamiltonian Matrices:
{\rm Let $H$ be Hamiltonian with representation $b(1\otimes j) +
p\otimes 1 + q\otimes i + r\otimes k$. Then,
\begin{itemize}
\item $H$ has a quadratic minimal polynomial, which equals 
$p(x) = x^{2} - \omega$,
with $\omega =  -b^{2} - p.p + q.q + r.r$, iff
$p.r = p.q = 0$ and $r\times q = -bp$. Notice, if $b\neq 0$, then the first
two conditions are subsumed by the last condition.
\item $H$ has a cubic minimal polynomial, which equals $p(x)
= x^{3} - (\omega + 2k)x$, with $\omega$ as in the quadratic minimal
polynomial case 
and $k$ as specified below,  iff one of
the following five mutually exclusive
conditions hold [See Remark (\ref{specialcasesi}), below, for special cases
of these conditions].  
\begin{enumerate}
\item $b\neq 0$ and the matrix $G = X^{T}X$, with
$X = [p\mid q \mid r]$, has the the matrix
\[
Y = \left (\begin{array}{ccc}
b^{2} + k & -p.q & -p.r\\
-p.q & r.r - k & -q.r\\
-p.r & -q.r & q.q - k
\end{array} \right )\]
with $k = \frac{1}{b}[p.(q\times r) - b(p.p)]$,
as its ``near inverse". Specifically $GY = b[p.(q\times r)]I_{3}$,
In this case the coefficient  $k$ in the given cubic
minimal polynomial is $\frac{1}{b}[p.(q\times r) -bp.p]$. 

\item $b = 0$, $r\times q\neq 0$, $p=0$ $r.q = 0$ and $q.q =
r.r$. In this case $k = r.r + q.q$

\item $b = 0$, $r\times q\neq 0$, $p\neq 0$, $p. (q\times r)
= 0$, $p.q = 0$, $r.q=0$, $p.r\neq 0$ and
$(r.r)^{2} + (p.r)^{2} =
(q.q)(r.r)$. In this case $k = r.r$.
 
\item $b = 0$, $r\times q\neq 0$, $p\neq 0$, 
$p.r = 0$, $q.r = 0$, $p.q\neq 0$, and
$(q.q)^{2} + (p.q)^{2} = (q.q)(r.r)$.  In this case, $k = q.q$.

\item $b=0, r\times q\neq 0, p\neq 0$, $p. (q\times r) = 0$,
$p.q\neq 0 \neq p.r$ and
the following four quantities are all equal to  
$-\frac{q.p}{q.r}(p.r)$,
\[
\mid\mid q\mid\mid^{2} +
\frac{p.r}{p.q}(r.q), \mid\mid r\mid\mid^{2} +
\frac{p.q}{p.r}(r.q), \mid\mid r\mid\mid^{2} - \frac{ [(q.p)^{2} + (q.r)^{2}]}
{\mid\mid q\mid\mid^{2}},
\mid\mid q\mid\mid^{2} 
-\frac{ [(r.p)^{2} + (q.r)^{2}]}{\mid\mid r\mid\mid^{2}}
\]
In this case $k = -\frac{q.p}{q.r}(p.r)$.
\end{enumerate}

\item If none of the above conditions hold, the minimal
polynomial is the characteristic polynomial, which equals  
 \[
p(x) =
x^{4} - 2\omega x^{2}
- (4b^{2}\mid\mid p \mid\mid^{2} + 8b p.(r\times q)
+ 4\mid\mid r\times q\mid\mid^{2} -\omega^{2} - 4(p.q)^{2}
- 4(p.r)^{2}) 
\] 
\end{itemize}} 

\end{theorem}

\begin{theorem}
\label{perskewsymm}
Minimal Polynomials of Perskewsymmetric Matrices:
{\rm Let $P$ be a perskewsymmetric matrix with representation
$r\otimes i + j\otimes s + \alpha (1\otimes i) + \beta (j\otimes 1)$.
Then,
\begin{itemize}
\item $P$ has a quadratic minimal polynomial iff one of the following three
mutually exclusive sets of conditions hold. These are: i)$\alpha = 0,
\beta \neq 0, s =0$; ii)$\beta = 0, \alpha\neq 0, r = 0$;
iii) $\alpha = \beta = 0$ and either $r\times j = 0$ or
$s\times i = 0$. In each of these cases the minimal polynomial
is $x^{2} - \lambda^{2}$, with $\lambda^{2} = r.r + s.s -\alpha^{2}
-\beta^{2}$.
\item $P$ has a cubic minimal polynomial iff $\alpha^{2} - \beta^{2}
= s.s - r.r$ (without any of the conditions in the quadratic
minimal polynomial case occurring). In this case the minimal polynomial
is $x^{3} - (\lambda^{2} +2\alpha^{2} - 2 (s.s))x$.
\item If none of the above conditions hold
the minimal polynomial is the characteristic polynomial which equals 
\[
p(x) =
x^{4} -2\lambda^{2}x^{2} -
[4\beta (s.s) - 4\alpha (r.r) + 4\alpha^{2}\beta^{2} - 4\mid\mid r\times j
\mid\mid^{2}.\mid\mid s\times i \mid\mid^{2} - \lambda^{4}]
\] 
\end{itemize}
}           

\end{theorem}

\noindent {\bf Sketch of the Proof:} We will illustrate the 
calculations involved by proving the conditions 
for quadratic and  cubic minimal polynomials for a Hamiltonian matrix $H$.

\noindent {\it Quadratic Case:} $H^{2}$'s quaternionic representation is
\[
H^{2} = (-b^{2} - p.p + q.q + r.r) (1\otimes 1)
+ 2 (r\times q + bp)\otimes j - (2p.q)1\otimes j - (2p.r)1\otimes k
\]
According to Proposition (\ref{shortlist}) if at all $H^{2}$ is linearly
dependent on a lower power of $H$, that power has to be $1\otimes 1$.
A necessary and sufficient condition for that to happen is evidently
\[
r\times q = -bp, p.q = 0, p.r = 0
\] 
Clearly, if $b\neq 0$, this set of conditions is equivalent to
$bp = q\times r$. If these conditions hold the minimal polynomial of $H$
is $x^{2} - \omega$, with
\[
\omega = -b^{2} - p.p + q.q + r.r
\]

\vspace*{1mm}

\noindent {\it Cubic Case:}   
By a direct calculation, which makes copious use of the vector triple
identity [Equation (\ref{vectortriple})], 
one finds that 
\begin{eqnarray*} 
H^{3} & = &\omega  H + 2[p. (q\times r) - bp.p]1\otimes j 
+ 2[-b^{2}p + b(q\times r) + (p.q)q + (p.r)r]\otimes 1\\
+ & & 2[-(p.q)p -b(r\times p) + (r.r)q - (r.q)r]\otimes i\\ 
+ &  &2[-(p.r)p - b(p\times q) + (q.q)r - (q.r)q)]\otimes k
\end{eqnarray*}
 
In view of Proposition (\ref{shortlist}),  for $H$ to have a cubic
minimal polynomial, therefore there has to be a real $k$ such
that 
\begin{equation}
\label{seeminglyadditional}
[p. (q\times r) - bp.p] = kb
\end{equation}
and further that
\begin{eqnarray}
\label{one}
(b^{2} + k)p - (p.q)q - (p.r)r & = & b(q\times r) \\ \nonumber
-(p.q)p + (r.r - k)q - (r.q)r & = & b(r\times p) \\ \nonumber
-(p.r)p - (q.r) q + (q.q - k)r & = & b(p\times q)
\end{eqnarray}
 When this happens, in absence of the conditions for a quadratic
minimal polynomial, the minimal polynomial of $H$ is
\[
p(x) = x^{3} - (\omega + 2k)x
\] 
There are now two possibilities. 
\begin{itemize}
\item $b\neq 0$, or
\item  $b = 0$.
\end{itemize}

In the former case, we find 
\[
k = \frac{1}{b}[p.(q\times r) -bp.p]
\]
Next, noting that $G$ is the Gram matrix of $X =
[p\mid q \mid r]$, one finds that taking the inner product
on both sides of Equation (\ref{one}) successively with $p,q,r$
yields \[
GY = b [p.(q\times r)]I_{3}
\] 
Conversely if $GY = b [p.(q\times r)]I_{3}$, one easily obtains Equation
(\ref{one}). For instance, if we denote by $v$ the vector
$(b^{2} + k)p - (p.q)q - (p.r)r$, then, in view of the
first column of $Y$, we find $v.r = 0 = v.q$, and hence $v$ is
proportional to $q\times r$. The remaining entry in this column
of $Y$ confirms that $v$ is indeed $b (q\times r)$. Hence $H$ has the
stated minimal polynomial.

Now suppose, $b = 0$. Then we first need $r\times q\neq 0$, for
otherwise we would have 
\[
2 (bp + (r\times q)) = 0
\] which corresponds
to the quadratic minimal polynomial case.
Further, under the condition $b = 0$, Equations (\ref{seeminglyadditional})
and (\ref{one}) reduce to
\begin{equation}
\label{seeminglyadditionalII}
p. (q\times r) = 0 
\end{equation}
and 
\begin{eqnarray}
\label{oneII}
 kp - (p.q)q - (p.r)r & = & 0 \\ \nonumber
-(p.q)p + (r.r - k)q - (r.q)r & = & 0 \\ \nonumber
-(p.r)p - (q.r) q + (q.q - k)r & = & 0 
\end{eqnarray}
   
Now the analysis
of the conditions equivalent to $H$ having the
stated minimal polynomial may be divided into two further cases: 
\begin{itemize}
\item $p=0$, or
\item $p\neq 0$.
\end{itemize}

Suppose first that $p$ is zero. Then Equation (\ref{seeminglyadditionalII})
and the first Equation in the system (\ref{oneII}) are trivially
satisfied, while the remaining two equations of Equation
(\ref{oneII}) yield
\begin{eqnarray*}
(r.r-k)q & = & (r.q)r\\
(q.q-k)r & = & (q.r)q 
\end{eqnarray*}
These two equations contradict $r\times q\neq 0$, unless
$r.r = q.q = k$ and $q.r = 0$. Conversely these two conditions are trivially
sufficient to ensure that $H$ has the said cubic minimal
polynomial when $p = 0 = b$. 

Next, suppose $p\neq 0$. Then certainly the linear independence of
$q$ and $r$,  and the linear dependence of $p$ on them is required.
Now at least one of $p.q$ or $p.r$ is not zero,
for otherwise $p$ becomes zero, contradicting the starting assumption
for this case.

Now the analysis may be divided into three cases:
\begin{itemize}
\item $p.q =0$, but $p.r \neq 0$. 
Then taking the inner product of the first equation in (\ref{oneII})
with $q$ forces $r.q = 0$. 
Next, by taking the inner product of 
the same equation with $r$ yields $k = r.r$. The second equation
also forces $k=r.r$. The third equation is trivially satisfied
upon taking inner product with $q$, while taking inner product with
$r$ forces $k = \frac{(q.q)(r.r) - (p.r)^{2}}{r.r}$.
Hence, we necessarily require $(r.r)^{2} + (p.r)^{2} =
(q.q)(r.r)$. Conversely, if these conditions hold, then the vectors
formed by the left hand sides of Equation (\ref{oneII}), which
are in the span of $q$ and $r$, are by construction orthogonal to $q$ and $r$.
Hence they must be zero. Thus, Equations (\ref{seeminglyadditionalII})
and (\ref{oneII}) are satisfied and hence $H$ has the stated cubic
minimal polynomial.
\item $p.r = 0$ and $p.q\neq 0$. Then, by an argument similar to the
one above, a necessary and sufficient set of conditions are given
by i) $q\times r \neq 0$; ii) $p.(q\times r) = 0$; iii)$r.q = 0$; iv)
$(q.q)^{2} + (p.q)^{2} = (q.q)(r.r)$.  In this case, $k = q.q$.

\item Neither $p.q$ nor $p.r$ is zero. Then, first by taking inner product
with $q$ of the last equation of the system of (\ref{oneII}), for instance,
one sees that $r.q\neq 0$. Next taking the inner product with respect to
$q$, first and then with respect to $r$ of all equations in
the system (\ref{oneII}), one arrives at six possible expressions for
$k$. Of these two are already equal to $-\frac{q.p}{q.r}(p.r)$.
The remaining four are
\[
\mid\mid q\mid\mid^{2} +
\frac{p.r}{p.q}(r.q), \mid\mid r\mid\mid^{2} +
\frac{p.q}{p.r}(r.q), \mid\mid r\mid\mid^{2} - \frac{ [(q.p)^{2} + (q.r)^{2}]}
{\mid\mid q\mid\mid^{2}},
\mid\mid q\mid\mid^{2} 
-\frac{ [(r.p)^{2} + (q.r)^{2}]}{\mid\mid r\mid\mid^{2}}
\]
Hence necessarily these four quantities are equal to each other and
to $-\frac{q.p}{q.r}(p.r)$.
Conversely, these conditions are sufficient to ensure that
$H$ has the said cubic minimal polynomial with $k
= -\frac{q.p}{q.r}(p.r)$.

\end{itemize}

$\diamondsuit$ 

\begin{remark}
\label{specialcasesi}
{\rm There are some special cases of the above result for
the stated cubic minimal polynomial
for a Hamiltonian matrix $H$, which deserve mention.

\begin{itemize}
\item  
First if $b\neq 0, p\neq 0$, and $p,q,r$ are collinear, then $H$ has the
given cubic minimal polynomial iff 
\[
b^{2} = p.p + q.q + r.r
\]
In this case $k = -p.p$. Indeed, in this case both the matrices
$G$ and $Y$ are rank one matrices, and the condition $GY = 0$
then is equivalent to $b^{2} = p.p + q.q + r.r$.
 
Note this contains the
special case that $q=r=0$. In this case, $H$ is also skew-symmetric,
and we find that a necessary and sufficient condition for $H$ to have
the given polynomial as its minimal polynomial is $p.p = b^{2}$.
This, as is easily seen, is in keeping with the 
conditions for a skew-symmetric
matrix to have a cubic minimal polynomial.

\item A second special case, diametrically opposed to the
previous one, occurs when the vectors $p,q,r$ are all non-zero, and satisfy
$q\times r = \alpha p, r\times p
=\beta q, p\times q = \gamma r$, for some non-zero real numbers
$\alpha , \beta , \gamma$. One then finds that $\alpha\neq b$
(for otherwise, we would have a quadratic minimal polynomial).
In this case $k = (\alpha - b)b$ and $G$ and $Y$ are both diagonal.
Then the condition $GY = bp. (q\times r)I_{3}$ is equivalent to
$\beta = \gamma$ (equivalently $q.q = r.r$) and
\[
b\beta^{2} + (\alpha - b)b - \alpha b^{2} = 0
\]
These conditions are satisfied if, for instance,
$b=\beta = \gamma$, $\alpha \neq b$ and $p.p = b^{2}$.

\item Note when $b= 0 = p$, $H$ is a symmetric, Hamiltonian matrix.
The conditions stated above for a cubic minimal polynomial for
$H$ also follow from Theorem (\ref{symmetric}) below.

\end{itemize}}  

\end{remark}

\begin{remark}
\label{assymetricroles}

{\rm Note that there is an asymmetry in the role of $p$
(vis a vis $q,r$) in the matrix $Y$ intervening in
the conditions for a cubic minimal polynomial for a Hamiltonian matrix $H$. 
This is not surprising since
$p$ stems from the anti-symmetric part of $H$, while $q, r$ stem from
the symmetric part of $H$.} 

\end{remark}

Next we study minimal polynomials for skew-Hamiltonian and symmetric
matrices. Now Proposition (\ref{shortlist}) does not apply. Nevertheless
we will find that the former always have quadratic minimal polynomials,
and this is an illustration of the utility of quaternions.
For the latter, in order to minimize bookkeeping, we suppose
they are traceless. Once the minimal polynomial of these are found, those
of symmetric matrices with non-zero trace are easily found.

\begin{theorem}
\label{skewhamil}
Minimal Polynomials of Skew-Hamiltonian Matrices:
{\rm Let $W$ be skew-Hamiltonian with representation $b(1\otimes 1)
+ p\otimes j + 1\otimes (ci + dk)$. Then $W$ has a quadratic
minimal polynomial, which equals $p(x) = x^{2} - 2bx + \kappa$, with
$\kappa = b^{2} - \mid\mid p \mid\mid^{2} + c^{2} + d^{2}$.
.}
   
\end{theorem}

\begin{theorem}
\label{symmetric}
Minimal Polynomials of Symmetric Matrices:
{\rm Let $S$ be a non-zero,  traceless symmetric matrix with representation
$p\otimes i + q\otimes j + r\otimes k$. Then
\begin{itemize}
\item i) $S$ has the quadratic minimal polynomial $p(x) = x^{2}-\lambda^{2}$
iff the rank of $X = [p, q, r]$ is one. 
In this case $\lambda^{2} = (p.p + q.q + r.r)$.
\item ii) $S$ has the quadratic minimal polynomial $p(x) = x^{2}
-2lx - \lambda^{2}$ iff $p\times q = lr, q\times r = lp, r\times p = lq$,
for the same non-zero $l$ and $\lambda\neq 0$, as in i) above.   
\item $S$ has the cubic minimal polynomial
$p(x) = x^{3} - (\lambda^{2} + 2\alpha)x$ iff the rank of $X$ is two
and one of the three following mutually 
exclusive conditions hold:
\begin{itemize}
\item Upto cyclic permutations of $p,q,r$,   
$p.q = 0,  r\times p = 0 = q\times r, p.p = q.q$.
In this case
$\alpha = p.p$.
\item Upto cyclic permutations of $p, q$ and $r$, one has 
$p\times q = 0, r\times p\neq 0, q.r = 0, p.p + q.q = r.r$.
In this case $\alpha = r.r$ (note the conditions $p\times q = 0$
and $q.r = 0$ imply $r.p = 0$ also).

\item  None of $p\times q, q\times r$ or $r\times p$ is zero and the
following set of equalities holds
\[
\mid\mid r\mid\mid^{2} - \frac{[(r.p)(r.q)]}{p.q}
= \mid\mid q \mid\mid^{2} - \frac{[(q.p)(r.q)]}{p.r}
= \mid\mid p \mid\mid^{2} - \frac{[(r.p)(p.q)]}{q.r}
\]
In this case $\alpha = \mid\mid p \mid\mid^{2} - \frac{[(r.p)(p.q)]}{q.r}$.
 
\end{itemize}   
\item When the degree of the minimal polynomial
is four, the minimal (and characteristic) polynomial is $p(x) = x^{4}
-2\lambda^{2}x^{2} - 8(p.(q\times r))x + [\lambda^{4} -
4 (\mid\mid q\times r\mid\mid^{2} + \mid\mid r\times p\mid\mid^{2}
+ \mid\mid p\times q\mid\mid^{2}]$.    
\end{itemize}
}
\end{theorem}

\noindent {\it Note:}
In the case of symmetric matrices, there are other cubic minimal
polynomials. Expressions and conditions for them can be found, but they
do not have elegant geometric interpretations, and so we omit them.
 
\noindent {\it Sketch of the proof:} Once again we illustrate the
quadratic and cubic minimal polynomial case for traceless, symmetric matrices.
One first finds that $S^{2}$
is given by
\[
S^{2} = (\mid\mid p\mid\mid^{2}
+ \mid\mid q\mid\mid^{2} + \mid\mid r\mid\mid^{2})
1\otimes 1 + 2 (p\times q)\otimes k + 2 (q\times r)\otimes i
+ 2 (r\times p)\otimes j
\]
Clearly then $p(x) = x^{2} - \lambda^{2}$ can annihilate
$S$ iff $p\times q = q\times r = r\times p = 0$, i.e.,
iff $[p, q, r]$ has rank one. When this holds
$\lambda^{2} = p.p + q.q + r.r$.

Similarly, $p(x) = x^{2} - 2lx - \lambda^{2}$ annihilates $S$ iff
$p\times q = lr, q\times r = lp, r\times p = lq$ for the same
non-zero $l$. When this happens $\lambda^{2}$ has to be necessarily
$p.p + q.q + r.r$.   

Next a calculation shows that
\begin{eqnarray*}
S^{3} & = & \lambda^{2}S + 6p.(q\times r) 1\otimes 1\\ 
 & \ & + 2[q\times (p\times q) - r\times(r\times p)\otimes i\\
& \ & + 2[r\times (q\times r) -p\times (p\times q)]\otimes j
+ 2[p\times (r\times p) -q\times (q\times r)]\otimes k
\end{eqnarray*}
It follows that for $S$ to have the desired
minimal polynomial one needs $p.(q\times r) = 0$ and 
that the following condition,{\it and all cyclic permutations of it}, have to
hold
\begin{equation}
\label{tripleidentity}
(q\times p )\times q + (r\times p)\times r = \alpha p
\end{equation}

for the  {\it same} non-zero $\alpha$. 

The condition $p.(q\times r) = 0$
forces the rank of $X = [p,q,r]$ to be atmost two. It has to be two,
since the rank one case corresponds to a quadratic minimal polynomial.
Hence rank of $X^{T}X = 2$. Since $X^{T}X$ is positive semidefinite,
at least one principal minor of order two has to be non-zero. 
Hence further analysis can be divided into three mutually exclusive cases:

\begin{itemize}
\item Precisely one  $2\times 2$
principal minor of $X^{T}X$ is non-zero - say the one
corresponding to the pair $(p,q)$. Thus $r\times p = 0  =q\times p$,
but $p\times q\neq 0$. So the system (\ref{tripleidentity}) reduces
to \[
q\times (p\times q) = \alpha p;(p\times q)\times p
= \alpha q; \alpha r = 0
\] Hence, the linear independence of $p,q$ first forces
$p.q = 0$ and $q.q = p.p = \alpha$. This implies $\alpha\neq 0$ and
hence $r=0$. Conversely these conditions
are sufficient for $S$ to have the stated minimum polynomial.

\item Precisely two of the $2\times 2$
principal minors of $X^{T}X$ are zero, say those 
corresponding
to the pairs $(r,p)$ and $(q,r)$. In particular, $p\times q = 0$.
Writing out the system
(\ref{tripleidentity}) under these assumptions, we  find
that \[
r.p = 0, \alpha = r.r, r.q =0, \alpha = p.p + q.q 
\] 
So the stated conditions are necessary and it is easy to see their sufficiency
as well.

\item None of the $2\times 2$ principal minors of $X^{T}X$ are zero.
Thus each of the pairs $(p,q)$, $(q,r)$ and $(r,p)$ are linearly
independent, but each of the three vectors is linearly dependent
on the remaining two. Then the system (\ref{tripleidentity})
is equivalent to
\begin{eqnarray} 
\label{newsystemsymmi}
(q.q + r.r - \alpha )p & = & (q.p)q + (r.p)r\\ \nonumber
(r.r + p.p - \alpha )q & = & (q.r)r + (q.p)p\\ \nonumber
(p.p + q.q - \alpha )r & = & (r.p)p + (r.q)q
\end{eqnarray}

Then the first equation in the last system says $q.p\neq 0$. Indeed,
if $q.p = 0$, then the linear independence of the pair of vectors $(r,p)$
would force $r.p = 0$ as well. But then, $p$ being linearly
dependent on the pair $(q,r)$ would have to
be zero. Similarly $r.p\neq 0$ and $q.r\neq 0$.

Now  successively
taking the inner product of the first equation in the above system
with $q, r$; of the second equation with $r,p$; and the third
equation with $p,q$, yields six possible expressions for $\alpha$.
Of these three are identical. Thus, we have three distinct expressions
for $\alpha$, which have therefore got to coincide, i.e., it is
necessary that
\[
r.r - \frac{[(r.p)(r.q)]}{p.q}
= q.q - \frac{[(q.p)(r.q)]}{p.r}
= p.p - \frac{[(r.p)(q.p)]}{q.r}
\]
Conversely if the above equalities hold, then the vectors represented
by the left hand sides of the system (\ref{newsystemsymmi}) are 
equal to the corresponding right hand sides. Hence these conditions
are necessary and sufficient for $S$ to have the stated cubic minimal
polynomial.
\end{itemize}     
 
\begin{remark}
\label{neededforcubic}
{\rm 
It is not enough for $X = [p, q, r]$ to have rank $2$ for  a symmetric
$S$ to have the stated
cubic minimal polynomial. The remaining conditions are needed.
In fact, it turns out that all other cases
when $X$ has rank two correspond to fourth degree minimal polynomials.}
\end{remark}

We next consider matrices in $SO(4, R)$. II) of Proposition (\ref{shortlist})
applies to such matrices.

\begin{theorem}
Minimal Polynomials of Special Orthogonal Matrices:
{\rm Let $G\neq I, -I$ be represented by $u\otimes v$. Let $u_{0}
= {\mbox Re} (u), v_{0} = {\mbox Re}(v)$. 
Then,
\begin{itemize}
\item $G$ has minimal polynomial $x^{2} - 1$ iff $u_{0} = v_{0} = 0$.
\item $G$ has minimal polynomial $x^{2} + ax + 1$ iff either
(but not both) ${\mbox Im} (u) = 0$ or ${\mbox Im} (v) = 0$. In this
case, $a = -2v_{0}$ (resp. $-2u_{0}$).  
\item $G$ has minimal polynomial $x^{3} -ax^{2} + ax  -1$
iff $u_{0} = v_{0}\neq 0$. In this case $a = 4u_{0}v_{0} - 1$.
\item $G$ has minimal polynomial $x^{3} + ax^{2} + ax + 1$
iff $u_{0} = -v_{0}\neq 0$. In this case $a = -(1+4u_{0}v_{0})$.
\item If none of the above conditions hold, $G$'s minimal polynomial
is its characteristic polynomial which equals $x^{4} + ax^{2}
+ bx^{2} + ax + 1$, with $a= -4u_{0}v_{0}, b = 4u_{0}^{2} + 4v_{0}^{2} - 2$.
 \end{itemize}
}
\end{theorem}

{\it Sketch of the proof:} First, since $G$ is neither
$I$ nor $-I$, ${\mbox Im}(u)$  and ${\mbox Im}(v)$ cannot be
simultaneously zero.

Next, using  $u^{2} = (2u_{0}^{2} -1) + 2u_{0} {\mbox Im}
(u)$ (and a similar expression for $v^{2}$), we see $G^{2}$ is represented
by
\[
(2u_{0}^{2} - 1) (2v_{0}^{2} - 1) (1\otimes 1) 
+ 2u_{0}(2v_{0}^{2} - 1)( {\mbox Im}(u)\otimes 1)
+  2v_{0}(2u_{0}^{2} - 1)(1\otimes  {\mbox Im}(v))  
+ 4u_{0}v_{0} ({\mbox Im}(u)\otimes {\mbox Im}(v))
\]
In view of Proposition (\ref{shortlist}) the only possible
candidates for a quadratic minimal polynomial are $p(x) = x^{2} -1$
and $p(x) = x^{2} + ax + 1$.

For $G^{2} = I$, it is necessary and sufficient that all of 
the following to hold:
\begin{itemize}
\item $(2u_{0}^{2} -1)(2v_{0}^{2} -1) = 1$.
\item $2v_{0} (2u_{0}^{2}-1) = 0$ or ${\mbox Im}(v) = 0$.
\item $2u_{0} (2v_{0}^{2}-1) = 0$ or ${\mbox Im}(u) = 0$ 
\item $4u_{0}v_{0} = 0$ or ${\mbox Im}(u) = 0$ or
${\mbox Im}(v) = 0$.
\end{itemize}

Suppose, ${\mbox Im}(u)=0$. Then
$u_{0}^{2} = 1$ and ${\mbox Im}(v)\neq 0$. 
So the second condition above forces $v_{0} = 0$.
But then the first condition above is not satisfied. Similarly
the condition ${\mbox Im}(v) = 0$ is untenable. Hence, 
${\mbox Im}(u)\neq 0\neq {\mbox Im}(v)$. Now the fourth and the
first conditions together force $u_{0} = 0 = v_{0}$.
Conversely, when $u_{0} = 0 = v_{0}$ we see, from    
conditions above, that $G^{2} = I$.
  
Next, consider  $G^{2} +aG +I = 0$. Suppose neither
${\mbox Im}(u)$ nor ${\mbox Im}(v)$ is zero. Then, by 
comparing the ${\mbox Im}(u)\otimes {\mbox Im}(v)$ coefficient,
it is necessary  that 
$a=-4u_{0}v_{0}$. 
But if $a=-4u_{0}v_{0}$, then (by comparing the $1\otimes 1$
coefficient)
we see $u_{0}^{2} + v_{0}^{2} =1$, while by comparing the coefficients
of ${\mbox Im}(u)\otimes 1$ and $1\otimes {\mbox Im}(v)$, we find
$u_{0}=0=v_{0}$. Hence, necessarily precisely one of ${\mbox Im}(u)$ or
${\mbox Im}(v)$  is zero. 

Suppose ${\mbox Im}(u) = 0$.
Then $u_{0}$ is $+1$ or $-1$. By absorbing the negative coefficient in $v$,
we may suppose $u_{0} =1$. In this case a further necessary condition
is $a=-2v_{0}$. Similarly, if ${\mbox Im}(v)=0$, $a=-2u_{0}$ is required.
Conversely, these conditions are easily seen to be sufficient
for $p(x) = x^{2}+ax +1$ to be the minimal polynomial of $G$.

Next we study the necessary and sufficient conditions for 
$G$ to have cubic minimal polynomials.
First, we find that 
\begin{eqnarray*}
G^{3} & = & u_{0}v_{0}(16u_{0}^{2}v_{0}^{2} -12u_{0}^{2}-12v_{0}^{2}+9)
1\otimes 1\\
& \ & v_{0}[20u_{0}^{2}v_{0}^{2} -12u_{0}^{2}-6v_{0}^{2} + 3]
{\mbox Im}(u)\otimes 1\\
& \ & u_{0}[20u_{0}^{2}v_{0}^{2} -12v_{0}^{2}-6u_{0}^{2} + 3]
1\otimes {\mbox Im}(v)\\
& \ & [16u_{0}^{2}v_{0}^{2} -4u_{0}^{2}-4v_{0}^{2} + 1]
{\mbox Im}(u)\otimes {\mbox Im}(v)
\end{eqnarray*}  
By Proposition (\ref{shortlist}) it suffices to consider when
$G^{3}+aG^{2}+aG+I = 0$ or $G^{3}-aG^{2} + aG-I = 0$ for suitable
constants $a$.

Writing out the former we get
\[
f_{1}(a, u_{0}, v_{0})1\otimes 1 + f_{2}(a, u_{0}v_{0}){\mbox Im}(u)\otimes 1
+f_{3}(a, u_{0}v_{0})1\otimes {\mbox Im}(v)
+f_{4}(a, u_{0}v_{0}){\mbox Im}(u)\otimes {\mbox Im}(v)
 = 0 \]
for some polynomials $f_{i}, i=1, \ldots , 4$, whose explicit form
we omit for brevity. Necessarily ${\mbox Im}(u)\neq 0
\neq {\mbox Im}(v)$ (for otherwise we would be in the case of lower
degree minimal polynomials). Hence a necessary and sufficient
condition for $G^{3}+aG^{2}+aG+I = 0$ is $f_{i}(a, u_{0} , v_{0})
=0, i=, \ldots , 4$. These four equalities are equivalent to
$u_{0} = -v_{0}$. Clearly we need $u_{0} =
-v_{0}\neq 0$ to preclude $G^{2} = I$. In this case,
we also find $a = -(1+4u_{0}v_{0})$. Similarly $x^{3}-ax^{2}+ax-1$
is the minimal polynomial iff $u_{0}=v_{0}\neq 0$ and 
$a= 4u_{0}v_{0} - 1$.

\begin{remark}
{\rm While (II) of Proposition (\ref{shortlist}) applies to other 
groups of matrices such as symplectic matrices, finding quaternioninc
representations for them is quite arduous, and the formulae
for such representations are not nearly as succinct as those for
matrices in $SO(4, R)$. In \cite{noncompactportion},
a quaternionic representation for $Sp(4, R)$ was obtained.
In particular, this was used to find a closed form formula for the
characteristic polynomial of such matrices. Extending this to find
expressions for the minimal polynomial remains to be investigated.}
\end{remark}
   
\section{Illustrative Applications}
In this section we work out a few sample applications of the foregoing results.
The first application shows that Jordan structure of skew-Hamiltonian
matrices is determined completely by
its minimal polynomial plus a single rank calculation (which can be
performed in closed form). 
The second application works out the Cayley transform
of skew-Hamiltonian matrices. Finally, we show how the minimal polynomial
calculation of symmetric matrices can be used to determine the singular
values of $3\times 3$ real matrices.

\noindent {\bf Jordan Structure of Skew-Hamiltonian Matrices:}

\begin{proposition}
\label{Skewhamjordanstructure}
{\rm Let $W$ be a non-scalar, skew-Hamiltonian with quaternionic representation
$b(1\otimes 1) + p\otimes j + 1\otimes (ci + dk)$. 
Then $W$ is diagonalizable
iff $\mid\mid p\mid\mid^{2} \neq c^{2} + d^{2}$. 
The Jordan normal form of $W$ is either ${\mbox diag} (b+\mu, b+\mu,
b-\mu, b-\mu )$ or ${\mbox diag} (J_{2}(b), J_{2} (b))$. Here
$\mu = \sqrt{\mid\mid p\mid\mid^{2} - c^{2} - d^{2}}$, and $J_{2}(b)$
stands for the standard $2\times 2$ Jordan block with $b$ as the corresponding
eigenvalue. 
Finally the characteristic
polynomial of $W$ is $p(x) = x^{4} - 4bx^{3} + (6b^{2} - 2\mu^{2})x^{2}
+ (4b\mu^{2} - 4b^{3})x + b^{4} + \mu^{4}  -2\mu^{2}b^{2}$.}
\end{proposition}

\noindent {\it Proof:} First since $W$ is non-scalar,
the quantity $\theta^{2} = \mid\mid p\mid\mid^{2} + c^{2}
+d^{2}$ is non-zero. Per Theorem (\ref{skewhamil}), $W$ has minimal
polynomial $x^{2} - 2b x +\kappa $, where 
$\kappa = b^{2} - \mid\mid p \mid\mid^{2} + c^{2} + d^{2}$. This polynomial
has roots $(b+ \mu , b-\mu )$, which are distinct iff $\mu \neq 0$,
whence the first conclusion.
The algebraic multiplicity of both the roots, $b + \mu$ and $b-\mu$,
as roots of the characteristic polynomial has to be two each, for any
other configuration of algebraic multiplicities would not yield
${\mbox Tr}(W) = 4b$. This yields the stated characteristic polynomial. 
Note that when $\mu = 0$, the sole eigenvalue is $b$ with algebraic
multiplicity four.

Next, when $\mu\neq 0$, $W$ is diagonalizable and, in view of the algebraic
multiplicities mentioned above, the corresponding Jordan form
is ${\mbox diag} (b+\mu, b+\mu,
b-\mu, b-\mu )$.

When $\mu = 0$, $b$ is a two-fold root of the minimal polynomial. Hence the
size of the largest Jordan block corresponding to the sole eigenvalue, $b$,
has to be $2$. Thus, the remaining Jordan blocks are either a single
Jordan block of size 2 or two Jordan blocks each of size $1$.
To determine
which possibility occurs, recall that for an $n\times n$ matrix 
$W$ \cite{hhorni} 
\[
n_{i} = r_{i-1} - 2r_{i} + r_{i+1}, i = 0, 1, \ldots , n-1
\]
where $n_{i}$ stands for the number of Jordan blocks of size $i$ corresponding
to a given eigenvalue $\lambda$ of $W$, and $r_{k} =
{\mbox rank} (W - \lambda I)^{k}$, with the convention that $r_{n+1}
= r_{n} = n - \nu$, with $\nu$ being the algebraic multiplicity of 
$\lambda$ as a root of the characteristic polynomial.

Let $Y = W - bI$. In view of the only possibilities for the
Jordan form of $W$ (when $\mu = 0$), it is obvious that the rank
of $Y$ is either $1$ or $2$.
From this we see that $W$ has 2 Jordan blocks of size $2$ each iff
${\mbox rank} (Y)$ is $2$, while it has one Jordan block of
size $2$ and two Jordan blocks of size $1$ each iff ${\mbox rank} (Y)$ is 
$1$. We will now show that only the former possibility can occur.

This can be seen in a variety of ways. For instance, ${\mbox rk}(Y)
= {\mbox rk}(Y^{T}Y)$ and the latter has rank 2 precisely when 
at least one of its $2\times 2$ principal minors is non-zero.
We will now show that at least one
$2\times 2$ principal minor $M_{ij}$ of $Y^{T}Y$ has to be non-zero.

Since $Y = p\otimes j + 1\otimes (ci + dk)$, a simple calculation
yields
\begin{equation}
Y^{T}Y = \left ( \begin{array}{cccc} 
\theta^{2} -2cp_{3} + 2dp_{1} & 2cp_{2} & -2dp_{3} & 2dp_{2}\\
2cp_{2} & \theta^{2} + 2cp_{3} + 2dp_{1} &  2dp_{2} & 2dp_{3}-2cp_{1}\\
-2cp_{1}-2dp_{3} & 2dp_{2} & \theta^{2} -2dp_{1} + 2cp_{3} & -2cp_{2}\\
2dp_{2} & 2dp_{3} -2cp_{1} & -2cp_{2} & \theta^{2} -2dp_{1} -2cp_{3}
\end{array} \right)
\end{equation}
Here, as before, $\theta^{2} = \mid\mid p\mid\mid^{2} + c^{2} + d^{2}$.
We will now show that even $5$ of the principal minors being zero
leads to the contradiction $\theta^{2} = 0$.
 
Specifically suppose
\begin{itemize}
\item $(\theta^{2} + 2dp_{1})^{2} - 4c^{2}p_{3}^{2} - 4c^{2}p_{2}^{2} = 0$
($(1,2)$ minor).
\item $\theta^{4} - (2cp_{3}-2dp_{1})^{2} -
(2cp_{1}+2dp_{3})^{2} = 0$ ($(1,3)$ minor).  
\item $(\theta^{2} -2cp_{3})^{2} - 4d^{2}p_{1}^{2} - 4d^{2}p_{2}^{2} = 0$
($(1,4)$ minor).
\item $(\theta^{2} + 2cp_{3})^{2} - 4d^{2}p_{1}^{2} - 4d^{2}p_{2}^{2} = 0$
($(2,3)$ minor).
\item $(\theta^{2} - 2dp_{1})^{2} - 4c^{2}p_{3}^{2} - 4c^{2}p_{2}^{2} = 0$
($(3,4)$ minor).
\end{itemize}  

Now the above facts regarding the $(1,2)$ and $(3,4)$ minors are
equivalent to $dp_{1} = 0$ (since $\theta^{2}\neq 0$).
Similarly the facts about the $(1,4)$ and $(2,3)$ minors are
equivalent to $cp_{3} = 0$. Hence these last two facts used in the
$(1,4)$ and $(1,2)$ minor say $\theta^{4} = 4d^{2}p_{2}^{2} 
= 4c^{2}p_{2}^{2}$. This means $d$ and $c$ are non-zero.
Hence, necessarily $p_{1} = p_{3} = 0$. Using this last piece of information
in the fact concerning the $(1,3)$ minor gives $\theta^{4} = 0$
- a contradiction. Hence ${\mbox rk}(Y^{T}Y)= {\mbox rk}(Y) = 2$. 
  
\vspace*{3mm}

\noindent {\bf Cayley transform of skew-Hamiltonian matrices:}
The Cayley transform of matrices provides a relationship between
matrix Lie groups and their Lie algebras. It is interesting to compute
it even for matrices not belonging to a Lie algebra. 
We do this below for skew-Hamiltonian matrices.

Let $\psi_{C}(A) = (I-A)(I+A)^{-1}$ be the Cayley transform of
$A$ which is assumed to be $4\times 4$ skew-Hamiltonian. Since 
$\psi_{C}(A)$ is not defined if $-1$ is an eigenvalue of $A$, we
suppose that $b$ equals neither $-(1+\mu )$ nor $-1 + \mu$.
We know from the results above that this ensures that $-1$ is not
in the spectrum of $A$. 

Since $A$'s minimal polynomial is quadratic,
we know $\psi_{C}(A) = c_{0}I + c_{1}A$, with $c_{0}, c_{1}$
some constants. So we get
\[
c_{1}A^{2} + (c_{1} + c_{0} +1)A + (c_{0} - 1)I = 0
\]
and hence, in view of the minimal polynomial of $A$,
\[
(2bc_{1} + c_{1} + c_{0} +1)A + (c_{0} -\kappa c_{1} - 1)I = 0
\]

This leads to the following system of equations for $c_{0}, c_{1}$:  
\begin{eqnarray} 
c_{0} - \kappa c_{1} & = & 1\\ \nonumber
c_{0} + (2b + 1)c_{1} & = & -1
\end{eqnarray}
This yields 
$c_{0} = \frac{2b + 1 - \kappa}{2b + 1 + \kappa}$ and $c_{1}
= \frac{-2}{2b + 1 + \kappa}$.

Hence,
\[
\psi_{C}(A) =  \frac{2b + 1 - \kappa}{2b + 1 + \kappa}I
+  \frac{-2}{2b + 1 + \kappa}A
\]

\noindent {\bf Singular values of $3\times 3$ real matrices}
One can use the  geometric characterizations of minimal polynomials
of traceless $4\times 4$ real symmetric matrices to infer
information about the singular values of real $3\times 3$ matrices.
This follows from the results in \cite{ni}, wherein the eigenvalues
of the symmetric matrix $X = p\otimes i + q\otimes j + r\otimes k$
are related to
the singular values of the real matrix $ Y = [p \mid q \mid r]$. Thus,
if $\sigma_{1}\geq \sigma_{2}\geq\sigma_{3}$ are the singular values of
$Y$ and $\tau = {\mbox sgn\ det} Y$ ($\tau = 0$ if $Y$ is singular),
then the eigenvalues of $X$ are $\lambda_{1} = \sigma_{1} + \sigma_{2}
+\tau\sigma_{3}, \lambda_{2} = \sigma_{1} - \sigma_{2}
- \tau\sigma_{3}, \lambda_{3} =-\sigma_{1} + \sigma_{2}
- \tau\sigma_{3}, \lambda_{4} = -\sigma_{1} - \sigma_{2}
+ \tau\sigma_{3}$.

From this expression and the fact that $X$'s minimal polynomial has to have
distinct roots, one can infer the following relation between
$X$'s minimal polynomial, and therefore the corresponding geometric conditions
on $p,q,r$ stated in Theorem (\ref{symmetric}), and $Y$'s singular values:

\begin{itemize}
\item $\sigma_{2} = 0 = \sigma_{3}, \sigma_{1}\neq 0$iff $X$ has minimal 
polynomial $x^{2} - c^{2}$.
\item $\sigma_{1} = \sigma_{2}\neq 0$ and $\sigma_{3} = 0$ iff
$X$ has minimal polynomial $x^{3} + cx$.
\item $\sigma_{1} = \sigma_{2} =\sigma_{3}\neq 0$ iff $X$ has minimal
polynomial $x^{2} -2lx - \lambda^{2}$.
\end{itemize}

\begin{remark}
\label{moresvdI}
{\rm The above list only contains those statements regarding the
singular values of $Y= [p, q, r]$ corresponding to the list of
minimal polynomials in Theorem (\ref{symmetric}). One can also infer
the following statements regarding the singular values of $Y$ by invoking
the diagonalizability of $X$. Alternatively, the statements below about the
singular values of $Y$ can be used to augment the list of minimal
polynomials of $X$. The corresponding conditions on $p, q, r$ are too
cumbersome to state. 
Partly because of this, and partly since that would have been
contrary to the spirit of the paper, these minimal polynomials were not
presented in Theorem (\ref{symmetric}) (cf., the note, immediately
following the statement of Theorem (\ref{symmetric}) and
Remark (\ref{neededforcubic}).
\begin{itemize}    
\item If $Y$ has rank $2$ and $\sigma_{1} \neq \sigma_{2}$ then $X$ has a 
quartic minimal polynomial.
\item $Y$ has a cubic minimal polynomial other than $x^{3} + cx$
iff $\tau\neq 0$ and either i) $\sigma_{1} = \sigma_{2}\neq\sigma_{3}$.
In this case no eigenvalue of $X$ is zero; or
ii) $\sigma_{2} = \sigma_{3}\neq\sigma_{1}$. In this case
$X$ has a zero eigenvalue iff $\sigma_{1} = 2\sigma_{2}$;
  
\end{itemize} }
\end{remark}

\section{Extensions}
There are a few potential extensions of this work which we will
discuss in this section. 

One trivial way to extend the above results is to consider block diagonal
matrices, with each block $4\times 4$. The minimal polynomial of
such a matrix is the least common multiple
of the minimal polynomials of the individual
blocks. Thus, when each of these blocks belongs to any
of the classes of matrices considered here, one can find in closed
form their minimal polynomials.

A second extension is to apply the theory of Clifford Algebras
to calculate minimal polynomials, since each Clifford algebra
arises as a suitable matrix algebra. In this regard we mention the
interesting work of \cite{cliffminpolyi}, where a symbolic calculation of
the so-called real minimal polynomial is used to calculate
exponentials of matrices. This, however, does not take into account
the involutions of Clifford algebras, and thus the structure of the
matrix is not used in finding minimal polynomials. In particular,
there are no analogues of the geometric conditions on quaternions
in the previous sections.

To understand the crux of the  
differences between our work and that in \cite{cliffminpolyi},
it is useful to note
the three features of $H\otimes H$ which enable our approach :
\begin{itemize}
\item i) $H\otimes H$ has a basis in which every element squares to plus
or minus $1$. Furthermore, any two elements in this basis commute or
anti-commute.

\item  ii) The matrix analogue
of the natural conjugation on $H\otimes H$ is matrix
transposition.

\item iii) The multiplication in $H\otimes H$ is intimately related
to the geometry of vectors in $R^{3}$.
\end{itemize}
For Clifford algebras the first
feature goes through verbatim. The second feature's effect is somewhat
diluted, inasmuch as the natural involutions of the theory
of Clifford Algebras (Clifford conjugation and
reversion), \cite{pertii,portei}, have easy matrix theoretic 
interpretations only in certain cases.
Finally, the third feature is completely lost. In the work of
\cite{cliffminpolyi}, only the first feature is used. Hence the structural
(i.e., geometric) conditions in this work
on a matrix's $H\otimes H$ representation,
for it to have a specific minimal polynomial, have no analogues
in \cite{cliffminpolyi}.

As mentioned in the previous paragraph, the three enabling features for
the $H\otimes H$ isomoprhism of $M(4, R)$ are diluted for
Clifford algebra isomorphisms of matrix algebras.
Nevertheless,
there are two ways in which the theory of Clifford algebras can be used
for the purpose at hand. First, one can uncover more classes for
$4\times 4$ matrices whose minimal polynomials can be calculated,
and whose Jordan structure is akin to those of skew-Hamiltonian
matrices. This is achieved by first considering matrices in $M(4, R)$
as elements of suitable Clifford algebras and inspecting their behaviour
under Clifford conjugation and/or reversion, and then representing
such matrices via quaternions. In some cases the
$H\otimes H$ representations of these matrices enables a complete 
characterization of their minimal polynomial. Arguably, one would have not 
been lead to consider these
classes otherwise. Secondly, one can
use Clifford algebra representations of matrices of larger size to
give a partial characterization of their minimal polynomials.
Whilst a complete characterization of possible minimal polynomials for
such matrices
is the subject of future work, one can already say more than what would
be possible without using Clifford algebras.

Let us now explore the first extension.
To that end, note that there are two standard involutions in the 
theory of Clifford algebras
- {\bf reversion and Clifford conjugation.},
\cite{pertii,portei} These are both anti-automorphisms. 
The matrix versions of these two involutions are easy for two classes
of Clifford algebras. For $Cl(n, 0)$, reversion is Hermitian conjugation,
while for $Cl(0,n)$ Clifford conjugation is Hermitian conjugation.
Representing $Cl(p+1, q+1)$ as $M(2, Cl(p,q))$ (the algebra of $2\times
2$ matrices with entries in $Cl(p,q)$), it is known that  
Clifford conjugation 
is represented as follows
\[
\left (\begin{array}{cc}
A & B\\
C & D
\end{array}\right )^{CC} = \left (\begin{array}{cc}
D^{rev} & -B^{rev} \\
-C^{rev} & A^{rev} 
\end{array}\right ) 
\]
while reversion is
\[
\left (\begin{array}{cc}
A & B\\
C & D
\end{array}\right )^{rev} = \left (\begin{array}{cc}
D^{cc} & B^{cc}\\
C^{cc} & A^{cc} 
\end{array}\right ) 
\]

Here, and in the balance of this section, $Z^{cc}$ (respectively,
$Z^{rev}$) stands for the Clifford conjugation (respectively, reversion)
of a matrix (or its Clifford representation) $Z$.

Let us illustrate how this can be used to find minimal polynomials
for matrices stemming $Cl(2,2)$. 
Since $Cl(1,1)$ is
$M(2, R)$, it follows that $Cl(2,2)$ is $M(4, R)$. 
On $Cl(1,1)$ reversion sends
$X$ to $R_{2}X^{T}R_{2}$ (which, in the notation introduced
in Section 2, is $X_{F}$), while 
Clifford conjugation sends $X$ to $-J_{2}X^{T}J_{2}$, which 
is $X_{H}$. Equivalently, since $X$ is $2\times 2$,
$X^{CC}$ is ${\mbox adj}(X)$, where, as usual, ${\mbox adj}(X)$ is
the classical adjugate of $X$. Thus, on $Cl(2,2)$ we get
\[
\left (\begin{array}{cc}
A & B\\
C & D
\end{array}\right )^{rev} = \left (\begin{array}{cc}
D_{H} & B_{H}\\
C_{H} & A_{H} 
\end{array}\right ) 
\]
Thus, if $X\in Cl(2,2)$ equals its own reversion, then
$A$ and $D$ are each other's adjugates, while $B$ and $C$ are
$2\times 2$ skew-Hamiltonian. A basis of $1$-vectors for $Cl(2,2)$ consists
of the following four matrices (written in $H\otimes H$ form):
\begin{equation}
f_{1} = -j\otimes k, f_{2} =i\otimes k, f_{3} =-1\otimes i,
f_{4} = -1\otimes j
\end{equation} 
This yields expressions for $2$-vectors etc.,
which we omit. Now, since $X$ equals its own reverse, it must be
a linear combination of the identity, $1$-vectors and $4$-vectors.
The last equation, for a basis of one-vectors for $Cl(2,2)$, thus
yields the following 
$H\otimes H$ representation of 
of the most general $X\in Cl(2,2)$ satisfying $X^{rev}=X$.
\begin{equation}
\label{Twotworev} 
a(1\otimes 1) + p\otimes k + 1\otimes s
\end{equation}
with $p, s$ pure-quaternions,
{\it with the latter having no $k$-component}. Thus such an $X$ is remarkably
similar to skew-Hamiltonian matrices, with the difference that the
{\it roles of $j$ and $k$ have been interchanged.}
Thus, we find, for instance that
$X^{2} = (p.p-s.s -a^{2})1\otimes 1 + 2aX$ and hence such an $X$'s
minimal polynomial is quadratic. We omit the similar
statements about the Jordan structure of such matrices, that this
minimal polynomial yields. If $X\in Cl(2,2)$ satisfies $X^{rev} = -X$,
then it has an $H\otimes H$ representation akin to that for a Hamiltonian
matrix, and therefore an analogue of Theorem
(\ref{Hamilt}) applies to it.

Similarly, if $X\in Cl(2,2)$ is minus its
own reversion, then i) $A=D^{F}$ and $B$ and $C$ are both
perskewsymmetric; and ii) the $H\otimes H$ representation of $X$ is
given by
\begin{equation}
\label{Twotwocliff}
X = a(1\otimes 1) + k\otimes p + q\otimes 1
\end{equation}
with $a\in {\mathcal R}, p,q\in {\mathcal P}$ and $q.k=0$.
Once again, this yields a quadratic minimal polynomial for $X$.

\begin{remark}
\begin{itemize}

\item {\rm A calculation shows that matrices represented by
Equation (\ref{Twotworev}) [resp. Equation (\ref{Twotwocliff})] 
are precisely those self-adjoint with respect to the non-degenerate
bilinear form on $R^{4}$ whose defining matrix is $M_{1\otimes k}$
(resp. $M_{k\otimes 1}$). Thus, the considerations of the previous paragraphs
yield a natural Clifford theoretic interpretations for these bilinear
forms.}
\item {\rm By passing to $Cl(3,1)$ and performing an analysis akin to
the one above for $Cl(2,2)$ one can show that those $X\in Cl(3,1)$
satisfying $X^{rev} = X$ are again given by Equation (\ref{Twotworev}),
while those satisfying $X^{CC} = X$ are skew-Hamiltonian matrices.}
\item {\rm It is worth emphasizing that the block structures of the matrices
considered in the previous paragraphs do not themselves reveal the
simplicity of their minimal polynomials. It is only by passing to
their $H\otimes H$ representations that we are lead to these results.}
\end{itemize}
\end{remark}
For higher dimensional matrix algebras arising from Clifford Algebras,
we do not (yet) have an exhaustive set of results. Nevertheless, 
some conclusions can be drawn, which would have been difficult
to arrive at without passing to Clifford Algebras. Let us illustrate
this via $Cl(0,6)$. This is $M(8, R)$. Furthermore, Clifford
conjugation is precisely matrix transposition in this case
and thus a matrix is anti-symmetric
iff it is minus its Clifford conjugation. Since Clifford conjugation
of a $p$-vector in $Cl(0,6)$ is minus itself iff $p=1,2,5,6$, an $8\times 8$
matrix is anti-symmetric iff it is a linear combination of of these
$p$-vectors. We use the following basis of $1$-vectors:
\begin{eqnarray}
e_{1} & = & \sigma_{z}\otimes\sigma_{x}\sigma_{z}\otimes I_{2}\\ \nonumber
e_{2} & = & \sigma_{z}\sigma_{x}\otimes I_{4}\\ \nonumber
e_{3} & = & \sigma_{x}\otimes\sigma_{z}\sigma_{x}\otimes\sigma_{x} \\ \nonumber
e_{4} & = & \sigma_{x}\otimes\sigma_{z}\sigma_{x}\otimes\sigma_{x}\\ \nonumber
e_{5} & = & \sigma_{x}\otimes I_{2}\otimes\sigma_{x}\sigma_{z}\\ \nonumber
e_{6} & = & \sigma_{z}\otimes\sigma_{x}\otimes\sigma_{z}\sigma_{x}
\end{eqnarray}
Here the $\sigma$'s are the usual Pauli matrices.
Using this one can write down a basis of $p$-vectors for $p=2,5,6$
which we omit for brevity. The typical $8\times 8$ anti-symmetric
matrix is thus a real linear combination
\begin{equation}
\label{genericcliffsix}
X = \sum_{i=1}^{6}p_{i}e_{i} + \sum_{i < j}p_{ij}e_{ij}
+ \sum_{i< j < k< l < m}p_{ijklm}e_{ijklm} + p_{123456}e_{123456}
\end{equation}
One can now list a set of mutually
exclusive conditions on these coefficients which are necessary
and sufficient for $X$ to have a quadratic minimal polynomial.
From Proposition (\ref{shortlist}) we know that the minimal polynomial
has to have the form $p(x) = x^{2} - \lambda^{2}$. Due to the more complicated
structure of Clifford multiplication on $Cl(0, 6)$ this list of conditions,
even for the quadratic case, are far too long to enlist. Therefore, we will
just give sample instances of these conditions. 

To that end, it is first noted that 
this set contains conditions of two types. The first
consists of conditions which merely equate some of the coefficients,
$p_{J}, J\subseteq\{1,2,3,4,5,6\}$ in Equation (\ref{genericcliffsix})
to zero. The latter consist of more complicated algebraic relations
between the $p_{J}$. To understand the difference between the two,
it is first noted that a $p$-vector and a $q$-vector either
commute or anti-commute. Conditions of the first type arise precisely
when all the summands in Equation (\ref{genericcliffsix}) anti-commute.
Under these circumstances the minimal polynomial of $X$ is clearly quadratic.
The latter set of condition arises when there are some commuting
summands in Equation (\ref{genericcliffsix}). In this case the corresponding
coefficients have to satisfy certain relations to ensure that 
$p(x) = x^{2} - \lambda^{2}$ is the minimal polynomial of $X$. 
By carefully considering the commutation relations between the
$1,2,5$ and $6$-vectors in $Cl(0,6)$ one can arrive at the aforementioned
conditions. 

Enlisted below are instances, first of the first type
of conditions and then of the second type of conditions.
\begin{itemize}
\item i) $X = p_{i}e_{i} + \sum_{k <i}p_{ki}e_{ki}
\sum_{j >i}p_{ij}e_{ij}  
 + p_{\alpha\beta\gamma\delta\epsilon}
e_{\alpha\beta\gamma\delta\epsilon}$, with $i\notin 
\{\alpha,\beta,\gamma,\delta,\epsilon\}$.
\item ii) $X=p_{i}e_{i} 
+ p_{\alpha\beta\gamma\delta\epsilon}e_{\alpha\beta\gamma\delta\epsilon} 
+p_{123456}e_{123456}$, with
$i\notin\{\alpha,\beta,\gamma,\delta,\epsilon\}$.

\end{itemize}

Examples of the second type of conditions are
\begin{itemize}
\item $X = p_{1}e_{1} + p_{2}e_{2} + p_{13}e_{13} + p_{23}e_{23}$
with $p_{1}p_{23} = p_{2}p_{13}$.
\item $X = p_{1}e_{1} + p_{23}e_{23} + p_{45}e_{45} + p_{12345}e_{12345}$
with $p_{1} = p_{45}, p_{23} = p_{12345}$ and $p_{1}$ equal to $p_{23}$
upto sign.
\end{itemize}

Finally, in all the cases above $\lambda^{2}$ is the Euclidean length squared
of the vector of coefficients describing $X$.   

\vspace*{9mm}
    
\noindent
{\it Octonions and Quadratic Minimal Polynomials:}
One special class of $8\times 8$ matrices which always have quadratic
minimal polynomials can be obtained via octonions. Whilst the octonions
are not associative, one can attach two $8\times 8$ matrices,
$\omega (a), \theta (a)$ to an octonion $a$,
\cite{octonionmatrix}. The former describes the
effect on an octonion upon left multiplication by $a$, while the latter
does the same for right multiplication by $a$. To describe them express
the octonion $a$ by a pair of quaternions, $a = (a_{1}, a_{2}), a_{i}\in H,
i=1,2$, via the Cayley doubling procedure, \cite{pertii}. Then 

\[
\omega (a) = \left ( \begin{array}{cc}
M_{a_{1}\otimes 1} & -M_{1\otimes \bar{a_{2}}}I_{1,3}\\
M_{a_{2}\otimes 1}I_{1,3} & M_{1\otimes \bar{a_{1}}}  
     \end{array}
\right )
\]
and
\[
\theta (a) = \left ( \begin{array}{cc}
M_{1\otimes \bar{a_{1}}} & -M_{\bar{a_{2}}\otimes 1}\\
M_{a_{2}\otimes 1} & M_{1\otimes a_{1}}
\end{array}
\right )
\]
Here $I_{1,3} = {\mbox diag}(1,-1,-1,-1)$.

Then, as shown in \cite{octonionmatrix}, the alternating identities
yield
\begin{eqnarray}
\omega (a^{2}) & = & (\omega (a))^{2}\\ \nonumber
\theta ( a^{2}) & = & (\theta (a))^{2}
\end{eqnarray}
Now since any octonion $a$ satisfies $a^{2} - 2 {\mbox Re}(a)a
 + \mid a\mid^{2} = 0$, we see that the $8\times 8$ matrices
$\omega (a)$ and $\theta (a)$ have the quadratic polynomial
$p(x) = x^{2} -  2 {\mbox Re}(a)x + \mid a\mid^{2}$ (as long
as $a\neq 0$. Next, since the octonions are not associative
one cannot expect $\omega (ab)$ (resp. $\theta (ab)$) to equal the
product $\omega (a)\omega (b)$ (resp. $\theta (a)\theta (b))$. Nevertheless,
if $ab\neq 0$, $\omega (ab)$ (resp. $\theta (ba)$) is similar to
$\omega (a)\omega (b)$ (resp. $\theta (a)\theta (b)$), \cite{octonionmatrix}.
Thus, the matrix $\omega (a)\omega (b)$ (resp. $\theta (a)\theta (b)$) 
has a quadratic minimal polynomial, $p(x) = x^{2}
- 2{\mbox Re}(ab) x + \mid ab\mid^{2}
= x^{2} - 2 <a, \bar{b}>x + \mid a\mid^{2}\mid b\mid^{2}$
(resp. $q(x) = x^{2} - 2{\mbox Re}(ba) x + \mid ba\mid^{2}$).

This is significant since the 
structure of the matrices $\omega (a)\omega (b)$ 
(resp. $\theta (a)\theta (b)$) {\it is more complicated} than that
of $\omega (c)$ (resp. $\theta (c)$), for an octonion $c$.   

We end this section with a discussion of how the method of
\cite{grammoment} can be combined with those of this work
to compute minimal polynomials of $4\times 4$ matrices not covered above.
The same discussion will also reveal why the method of
\cite{grammoment} requires more computation than that proposed
here. 

We  first briefly recall the method of \cite{grammoment} for 
computing the minimal polynomial of a matrix $X$ of size $n\times n$.
One first associates to the sequence
$\{I, X, X^{2}, \ldots \}$ the matrices $G_{i}, i=1, \ldots , n$, where
$G_{i}$ is the Gram matrix of the set of
matrices $\{I, X, X^{2}, \ldots , X^{i}\}$ with respect to the
inner product $<Y, Z> = {\mbox Tr}(Y^{T}Z)$ (here, for brevity, all
matrices are assumed to be real). 
Thus, for instance
\[
G_{2} = \left (\begin{array}{ccc}
{\mbox Tr}(I) & {\mbox Tr}( X) & {\mbox Tr}(X^{2})\\
{\mbox Tr}(X^{T}) & {\mbox Tr}(X^{T}X) & {\mbox Tr}(X^{T}X^{2})\\
{\mbox Tr}((X^{T})^{2}) & {\mbox Tr}((X^{T})^{2}X) & 
{\mbox Tr} (X^{T})^{2}X^{2})
\end{array}
\right )
\]
The method then, in essence, consists of two steps:
\begin{itemize}
\item One computes the ranks of the $G_{i}$'s.
Then the degree of the minimal polynomial of $X$ is $r$ iff the first
$i$ for which the rank of $G_{i}$ is lower than $i+1$ is $r$.

\item  In this
case it is also known that the kernel of $G_{r}$ is of dimension one.
Furthermore, it is guaranteed that there is a vector in the
kernel of $G_{r}$ whose last coefficient is non-zero. Normalizing this
coefficient to one yields a vector $(a_{0},a_{1}, \ldots , a_{r-1}, 1)$ in
the kernel of $G_{r}$. This vector yields the minimal polynomial of $X$
to be $p(x) = x^{r} + \sum_{i=0}^{r-1}a_{i}x^{i}$.
\end{itemize}

Thus, this method requires two steps i) Calculating the $G_{i}$
and their ranks successively till one detects a drop in rank.
Thus, this step requires a requisite number of trace calculations
plus one's favourite method to compute ranks; ii) Computing a non-zero
vector in the kernel of $G_{r}$. 

The first step is amenable to the methods used in this work, since
to find the trace of a matrix being represented in quaternion (or Clifford
Algebra) form, one has to only find the coefficient of the $1\otimes 1$
term in the matrix. This rarely
requires the full quaternionic expansion of the matrix.
However, even for the classes of structured matrices considered here,
these calculations involve more than those required by our methods. 
We illustrate 
this issue via the case of $4\times 4$ real symmetric matrices.
To detect a quadratic minimal polynomial, our method requires finding
only $X^{2}$. However, to find $G_{3}$ and check if its rank is two,
one needs terms such as ${\mbox Tr}(X^{3})$. While, this does not require
the full calculation of $X^{3}$, it requires more than a calculation of
$X^{2}$, because one has to find the $1\otimes 1$ term in $X^{3}$.   
                      
Even when the ranks of the $G_{i}$ have been computed and the degree
of the minimal polynomial found, one has to still find a non-zero element
of the kernel of $G_{r}$. This is typically difficult to do in closed form,
whereas the methods used here do produce the minimal polynomials
(for the classes of matrices considered here) in closed form. 

\section{Conclusions}
In this work a complete characterization of the minimal polynomials of
several important classes of $4\times 4$ real matrices, including those
of interest in applications, was provided.
These were illustrated by relevant applications such as the determination
of the Jordan structure of $4\times 4$ skew-Hamiltonian matrices.
Extensions of these results via the usage of Clifford algebras was indicated.
In particular, classes of matrices were found whose block structures
bely their close similarity, vis a vis
minimal polynomials, to skew-Hamiltonian and Hamiltonian matrices.
Extensions of the preliminary results announced
here for $M_{8}(R)$ will be  the subject of future investigations.


\begin{thebibliography}{99}

\bibitem{cliffminpolyi} R. Ablamowicz, ``Matrix Exponential
Via Clifford Algebras" {\it J. Nonlinear Mathematical Physics},
{\bf 5}, 294-313, 1998.

\bibitem{noncompactportion} Y. Ansari $\&$ V. Ramakrishna,
`` On The Non-compact Portion of $Sp(4, R)$ Via Quaternions",
{\it J. Phys A: Math. Theor}, {\bf 41},
335203, 1-12, (2008).

\bibitem{goongi} G. Chen, D. Church, B. Englert, C. Henkel,
B. Rohnwedder, M. Scully $\&$ M. Zubairy,
{\it Quantum Computing Devices: Principles, Design and Analysis},
Chapman $\&$ Hall CRC Press, Boca Raton, (2006).

\bibitem{fourporti}
T. Constantinescu, V. Ramakrishna, N. Spears, L. R. Hunt, J. Tong,
I. Panahi, G. Kannan, D. L. MacFarlane, G. Evans, and M. P. Christensen,
``Composition methods for four-port couplers in
photonic integrated circuitry",
{\it Journal of Optical Society of America A}, {\bf 23}, 2919-2931,
(2006).

\bibitem{nii} H. Fassbender, D. Mackey $\&$ N. Mackey,
Hamilton and Jacobi Come Full Circle:
Jacobi Algorithms For Structured Hamiltonian
Eigenproblems",
{\it Linear Algebra $\&$ its Applications}, {\bf 332}, 37- 80, (2001).


\bibitem{haconi} D. Hacon, ``Jacobi's Method for Skew-Symmetric Matrices",
{\it SIAM J. Matrix Analysis },
{\bf 14}, 619 - 628, (1993).

\bibitem{hhorni} R. A. Horn $\&$ C. R. Johnson, {\it Matrix Aanlysis},
Cambridge University Press (1990).


\bibitem{grammoment} R. Horn $\&$ A. Lopatin,
``The Moment and Gram Matrices, Distinct Eigenvalues and Zeroes,
and Rational Criteria for Diagonalizability",
{\it Linear Algebra and Its Applications},
{\bf 299}, 153-163, (1999).


\bibitem{kyf} C. R. Johnson, T. Laffe $\&$ C. K. Li,
``Linear Transformations on $M_{n}(R)$ That Preserve the Ky Fan $k$-Norm and
a Remarkable Special Case When $(n,k) = (4,2)$,
{\it Linear and Multilinear
Algebra}, {\bf 23}, 285 - 298, (1988).

\bibitem{pertii} P. Lounesto, {\it Clifford Algebras and Spinors},
II edition, Cambridge University Press (2002).

\bibitem{ni} N. Mackey, ``Hamilton and Jacobi Meet Again - Quaternions and
the Eigenvalue Problem", {\it Siam J. Matrix  Analysis},
{\bf 16}, 421 - 435, (1995).

\bibitem{niii} D. Mackey, N. Mackey $\&$ S. Dunleavy,
`` Structure Preserving
Algorithms for Perplectic Eigenproblems", {\it Electronic Journal
of Linear Algebra}, {\bf 13}, 10 - 39, (2005).

\bibitem{portei} I. R. Porteous, {\it Clifford Algebras and the Classical
Groups}, Cambridge U Press, (2009).

\bibitem{expistruc} V. Ramakrishna $\&$ F. Costa,
 ``On the Exponential of Some Structured Matrices",
{\it J. Phys. A - Math $\&$ General},
{\bf  Vol 37}, 11613-11627, 2004.

\bibitem{expisufour} V. Ramakrishna$\&$
H. Zhou,  ``On the Exponential of Matrices in $su(4)$",
{\it J. Phys. A - Math $\&$ General}, {\bf 39}
(2006), 3021-3034.

\bibitem{selig} J. M. Selig, {\it Geometrical Foundations of Robotics},
World Scientific, Singapore, (2000). 



   

\bibitem{octonionmatrix} Y. Tian, ``Matrix Representations of Octonions and
Their Applications", {\it Advances in Applied Clifford Algebras}, {\bf 10},
61-90, (2000).
 \end{thebibliography}
\end{document}